\documentclass[a4paper,10pt]{article}
\usepackage{amsmath}
\usepackage{color}
\usepackage{amssymb}

\title{Free Fermions on causal sets.}
\author{Johan Noldus}

\begin{document}

\maketitle

\begin{abstract}
We construct a Dirac theory on causal sets; a key element in the construction being that the causet must be regarded as emergent in an appropriate sense too.  We further notice that mixed norm spaces appear in the construction allowing for negative norm particles and ``ghosts''.   This work extends the theory initiated in \cite{Johnston1,Johnston2}
\end{abstract}
\section{Introduction}
Retrieving continuum concepts from a pure discrete setting or visa versa is always tricky as one often has the tendency to forget that they must be natural from the discrete perspective too and in case of Fermi fields on causal sets; this problem has defied anybody up till now.  In contrast to the basic spacetime geometric objects one requires for the scalar field; the concepts of vierbein and Clifford bundle are mandatory for the continuum description.  Given the complete absence of such concepts for causal set theory, the best one may hope for is the existence of an object which has no counterpart in the continuum and replaces the aformentioned mathematical gadgets.  The Dirac operator is replaced by what we call a generating structure which allows one to construct the relevant Green's function akin to the methods in the scalar case.  As is standard for such enterprise, this section starts from fairly conventional ideas and then moves gradually in the discrete direction in which some equalities are highlighted and others are merely approximated.  I am aware I could have presented section two more formally but have chosen not to do so in order to allow the reader to see himself what is essential and what not.  Section three contains the general construction of Dirac theory and reveals the possibility of negative norm particles and ghosts which have to enter the prescription.  The last section finally gives an explicit computation of massless Dirac theory on the diamond for a particular generating structure and comments on possible different physical theories for distinct generating structures.   
\section{The relevant Fermionic Green's kernels.}
We start this section by retrieving some results and terminology from Lorentzian geometry.  The one parameter family of Dirac operators $D_{m} = i \gamma^{a}e_{a}^{\mu} \nabla_{\mu} - m$ in signature $(+---)$ with global Lorentz covariance and advanced Green's kernels $A_{m}(x,y)$ obeys
$$D_{m} A_{m}(x,y) = \frac{\delta^{n}(x,y)}{\sqrt{-g(y)}} \, 1.$$  Denoting by $G_{A,m}(x,y)$ the advanced solution to the Klein Gordon equation, one calculates that, with $$S_{m}(x,z) = - \int A_{m}(x,y)A_{-m}(y,z) \sqrt{-g(y)}d^{n}y$$
the following equality holds
$$(\Box^2 + m^2)S_{m}(x,z) = \frac{\delta^{n}(x,z)}{\sqrt{-g(z)}}\, 1$$
and since $S_{m}$ is advanced, $S_{m}(x,z) = G_{A,m}(x,z) \, 1$.  We define now two different automorphisms $\star$ and $\, \overline{a}\,$ by $c^{\star} = c, \overline{c} = \overline{c}$ and $ - (\gamma^{a})^{\star} = \gamma^{a} = - \overline{\gamma^{a}}$.  One notices then that $- A_{-m}^{\star}(x,y) = A_{m}(x,y)$ and $\overline{A}_{m}(x,y) = A_{m}(x,y)$; discretizing according to the causal set scheme then gives that 
$$\sum_{y} A_{m}(x,y)A^{\star}_{m}(y,z) = \rho \,1 \otimes G_{A,m}(x,z) $$ implying $\left[ A_{m}, A_{m}^{\star} \right] = 0$ and the reality condition that $\overline{A}_{m} = A_{m}$.  Denoting by $\, \widetilde{} \,$ the composition of $\, \overline{a} \,$ with the reversion, taking into account that $- D_{-m}G_{A,m}(x,y) = A_{m}(x,y)$, one arrives at $\widetilde{A}_{m}(x,y) = A_{m}(x,y)$.  Causality is then implemented at the discrete level by demanding that $A_{m}$ has support on the support of the incidence matrix union the diagonal and global Lorentz covariance $U = u \delta_{x,y}$, $\widetilde{u} u = 1$ implying that $u^{\star} = u$, translates as $$A_{m}' = U^{-1} A_{m} U.$$  Actually any transformation with the appropriate symmetries in the commutant of $G_{A,m}$ leads to a recalibration of $A_{m}$.  We could try to show that such transformations lead to unitarily equivalent theories.  We have that
$$A_{m}(x,y) = - mG_{A,m}(x,y) - i\gamma^{a}e_{a}^{\mu}\partial_{\mu}G_{A,m}(x,y)$$ and $e_{a}^{\mu}\partial_{\mu}G_{A,m}(x,y) = \left[ P_{a}(I)\right](x,y)$ is a polynomial expression in terms of the incidence matrix $I$ where $I(x,y) = 0$ unless $x \prec \star y$ in\footnote{$x \prec \star y$ if and only if $x \prec y$ and there is no $z$ with $x \prec z \prec y$.} which case it equals one.   Hence $$\sum_y  (- m G_{A,m}(x,y) - i \gamma^{a} \left[ P_{a}(I)\right](x,y))(- m G_{A,m}(y,z) + i \gamma^{b} \left[ P_{b}(I)\right](y,z)) = \rho G_{A,m}(x,z).$$  One notices that not enough information is present in the causal set itself to find a unique solution in this way; the problem being that too many expressions can fit these equations and that the dimension is put in by hand. \\* \\*
We now persue the viewpoint that the causal set itself is \emph{emergent} from a spinor perspective in the following sense: consider $\mathbb{C}l_{\mathbb{R}}(p,q)$ and the trace state $\textrm{Tr}$ on it; then $K \in \mathbb{C}l_{\mathbb{R}}(p,q) \otimes \mathbb{R}^{n \times n}$, $K = A_{a \ldots c}\gamma^{a}\ldots \gamma^{c}$ with $A_{a \ldots c }(x,y) \in \mathbb{R}$ having support on the union of the diagonal and the support of $I$, weakly generates $I$ if and only if 
$$\textrm{Tr}(KK^{\star}) = I,$$ while it strongly generates $I$ if and only if
$$KK^{\star} + K^{\star}K = 2 \,1 \otimes I.$$  We say that $K$ generates $I$ ultrastrongly if and only if $$KK^{\star} = I.$$
 In case of one Clifford generator, weakly generating structures are also strongly generating.  Take for example the causal set associated to 
$\left( \begin{array}{cc}
	0 & 1 \\*
	0 & 0
\end{array} \right)$ then $\mathbb{C}l_{\mathbb{R}}(1,0)$ accomodates for a strongly generating $K$:
$$K = \frac{1}{2}\left( \begin{array}{cc}
	1+ \beta & 1 - \beta \\*
	0 & 1 + \beta
\end{array} \right)  $$ since it obeys $KK^{\star} = \left( \begin{array}{cc}
	0 & 1 \\*
	0 & 0
\end{array} \right)  $ with $\beta^2 = 1$.  In $\mathbb{C}$,  no generating structure exists which is hopeful since it selects the correct signature (i.e. positive instead of imaginary mass).  More in general, $$K = \left( \begin{array}{ccc}
	1+ \beta & \frac{1}{4}\left(1 - \beta \right) & 0 \\*
	0 & 1 + \beta &  \frac{1}{4}\left(1 - \beta \right) \\*
	0 & 0 & 1 + \beta
\end{array} \right)$$ and $$ L = \left( \begin{array}{cccc}
	1+\beta &  \frac{1}{4}\left(1 - \beta \right) & 0 & 0 \\*
	0 & 1 + \beta &  \frac{1}{4}\left(1 - \beta \right) & 0 \\*
	0 & 0 & 1 + \beta &  \frac{1}{4}\left(1 - \beta \right) \\*
	0 & 0 & 0 & 1 + \beta
\end{array} \right)  $$ show that the one dimensional structure persists since 
$$KK^{\star} = \left( \begin{array}{ccc}
	0 & 1 & 0 \\*
	0 & 0 & 1  \\*
	0 & 0 & 0
\end{array} \right)  $$ and $$LL^{\star}= \left( \begin{array}{cccc}
	0 & 1 & 0 & 0 \\*
	0 & 0 & 1  & 0 \\*
	0 & 0 & 0 & 1 \\*
	0 & 0 & 0 & 0 
\end{array} \right).$$  The causet $$I = \left( \begin{array}{ccc}
	0 & 0 & 1 \\*
	0 & 0 & 1\\*
	0 & 0 & 0 \end{array} \right)$$  is one dimensional from the spinor structure as 
	$$K = \left( \begin{array}{ccc}
	1+\beta & 0 & - \frac{1}{2} \beta \\*
	0 & 1+\beta & - \frac{1}{2} \beta \\*
	0 & 0 & 1 + \beta \end{array} \right)$$ ultrastrongly generates $I$.   For the causal diamond, the reader can find an ultrastrongly generating\footnote{The following one is ultra strongly generating $$K = \frac{1}{2} \left( \begin{array}{cccc}
	1+\beta & 1-\beta & 1-\beta & 0 \\*
	0 & 1 + \beta & 0 & 1-\beta \\*
	0 & 0 & 1 + \beta & 1-\beta \\*
	0 & 0 & 0 & 1+\beta
	 \end{array} \right).$$ 
} $K$ associated to $\mathbb{C}l_{\mathbb{R}}(1,0)$ and it remains a question to find the need for higher dimensional algebra's.  Here, we introduce the notion of past and future distinguishing vertices; given a generating matrix $K$ for a causet $I$, a vertex $v$ is future distinguishing if and only if $K_{vx} \neq rK_{vy} + s\,1$ for all vertices $x,y$ such that $I_{vx} = I_{vy} = 1$ and real numbers $r,s$.  We say that $K$ is future distinguishing if and only if all vertices are; likewise, one has the notion of past distinguishing.  So far, we learned that one should look for ultrastrongly generating future and past distinguishing matrices $K$, given an incidence matrix $I$.  The causet (the ''hook'') defined by $p_1 \prec p_2 \prec p_3 \prec p_5$ and $p_1 \prec p_4 \prec p_5$ cannot be accomodated for by $\mathbb{C}l_{\mathbb{R}}(1,0)$ in this way\footnote{It can be accomodated ultrastrongly by $\mathbb{C}l_{\mathbb{R}}(1,0)$, but those matrices are not past and future distinguishing.}, but it can be by $\mathbb{C}l_{\mathbb{R}}(1,1)$.  While the past and future distinguishing condition forces one to go beyond one generator, it isn't obvious it will provide you with more than two generators.  For example, the causet given by $p_0 \prec p_i \prec p_n$ for $i = 1 \ldots n-1$ carries many future and past distinguishing ultra strong generating structures in $\mathbb{C}l_{\mathbb{R}}(1,1)$.  One would expect the ''hook'' to depend upon three generators and not two.  Even for the simple case of the two element causal set $p_1 \prec p_2$, one might expect two generators since it should feel a space and time dimension\footnote{In that case, a generating structure is given by $$K = \frac{1}{2} \left( \begin{array}{cc}
	\beta + \delta & \delta - \beta \\*
	0 & \beta + \delta
	 \end{array} \right)$$ with $\delta^2 = - \beta^2 = -1$ and $\beta \delta + \delta \beta = 0$.}.  There is another relationship coming from the continuum which is that $$K = - K^{\star}.$$  Moreover, if $K$ is generating $I$, so is $K^{\star} = \overline{K}, -K, -K^{\star}$ in which case the above relationship reduces the abundancy.   For the retarded case, we may use $K^{\dag}$\footnote{$\dag$ is the composition of $T$ and the involution on the Clifford algebra defined by $(\gamma^a)^{\dag} = \gamma^a$ since the adjoint of the integral operator with kernel $i \partial_{x^{\mu}}G_{R}(x,y)$ equals $-i \partial_{y^{\mu}}G_{R}(y,x)$ on the Hilbert space with as kernel adjoint $\alpha(x,y)^{\dag} = \overline{\alpha(y,x)}$.   Now, on flat Minkowski, $\partial_{x^{\mu}} G_{R}(x,y) = - \partial_{y^{\mu}}G_{R}(x,y)$ which is precisely what we need.  } since $$I^{T}(x,y) = \textrm{Tr}{(KK^{\star})}^{\dag}(x,y) = \textrm{Tr}(K^{\dag}(K^{\dag})^{\star})(x,y)$$ in accordance with the continuum theory.  This implies the use of indefinite metric spaces with as scalar product, in the previous examples, 

	$$\langle v | w \rangle = \left( 
\begin{array}{cc} \overline{v}_1 & \overline{v}_2
\end{array}
  \right)  
  \left( 
\begin{array}{cc} 1 & 0 \\* 0 & -1
\end{array}
  \right)
  \left(
\begin{array}{c} w_1 \\* w_2
\end{array}
  \right).$$  Denoting the latter matrix by $\delta$ and assuming that $\beta$ is given by 
  $$\beta =  \left( 
\begin{array}{cc} 0 & 1 \\* 1 & 0
\end{array}
  \right)  $$  
	one observes that $\delta \beta + \beta \delta = 0$ and $\beta^{\dag} = \delta \beta \delta = - \beta$ where the adjoint is defined with respect to the indefinite product.  That is, the adjoint acts like $\, \widetilde{}  \,$ on $\mathbb{C}l_{\mathbb{C}}(1,0)$ and the total representation space is a $2n$ dimensional complex space with a scalar product of signature $(n,n)$\footnote{In general, to implement $\dag$, we need to find a self adjoint matrix $\gamma$, $\gamma^2 = 1$ which commutes with all hermitian generators and anti-commutes with all anti-hermitian generators.  For example in $Cl_{\mathbb{R}}(1,1)$ with generators $\beta$ and $\delta$, one can choose a representation in which 
$$\beta =  \left( 
\begin{array}{cc} 0 & 1 \\* 1 & 0
\end{array}
  \right) $$ and
$$\delta =  \left( 
\begin{array}{cc} i & 0 \\* 0 & -i
\end{array}
  \right)$$ while $\gamma = \beta$.  For four dimensional Dirac theory $\gamma=\gamma^{0}$.}.  A spectral theorem for self adjoint matrices exists and one has real eigenvalues associated to $k$ eigenvectors of positive and negative norm respectively and to $n-k$ pairs of null vectors associated to any \emph{complex} eigenvalue.  That is, any \emph{non-degenerate}\footnote{Self adjoint operators $A$ with a degenerate eigenvalue polynomial might have no spectral decomposition as is shown in the following example.  Pick $$\Gamma = \left( 
\begin{array}{cc} 0 & 1 \\* 1 & 0
\end{array}
  \right) $$ and consider the operator
$$A = \left( 
\begin{array}{cc} \lambda & 1 \\* 0 & \lambda
\end{array}
  \right)$$ then in $\mathbb{C}^{2}$ with scalar product $v^{H}\Gamma w$, the adjoint of $A$ equals $\Gamma A^{H}\Gamma$ and an elementary computation reveals that this equals $A$.} self adjoint operator $A$ can be written as
	$$A = \sum_{r=1}^{k}\left( \lambda_r |v_r \rangle \langle v_r | + \mu_r |w_r \rangle \langle w_r |  \right) + \sum_{r=1}^{n-k}
	\left( c_r|m_r  \rangle \langle n_r | + \overline{c}_{r} |n_r \rangle \langle m_r | \right)$$ where $\langle v_{i}|v_{j}
\rangle = \delta_{ij} = - \langle w_{i}	| w_{j} \rangle$, $\langle n_i |  m_j \rangle = \delta_{ij}$ and all other scalar products vanish.  We construct $A_{m}$ from $K$ and $K^{\star}$ so that $\overline{A}_{m} = A_{m}$ and the product equality $A_{m}A_{m}^{\star} =\rho\, 1 \otimes G_{A,m}$ holds.  In the massless case, where $G_{A} = aI$, $a > 0$ it follows that
$$A_{0} = i\sqrt{\rho a}K$$ satisfying $$A_0 \, A_{0}^{\star} = \rho aI = \rho \, 1 \otimes G_{A}$$ given that $K + K^{\star} = 0$\footnote{One notices that our proposal also satisfies $- A_{0}^{\star} = A_{0}$.} for a generating structure $K$ of $-I$\footnote{That is $KK^{\star} = - I$ or $K^2 = I$.}.  Johnston \cite{Johnston2} proposes a mass expansion for the massive propagator
$$A_{m} = A_{0} + \overline{b}A_{0}^2 + \overline{b}^2 A_{0}^3 + \ldots = A_{0}(1 - \overline{b}A_{0})^{-1}$$ where $\overline{b}$ is mass assymetric.  Summarizing, we construct from a future and past distinguishing generating structure $K$ for $-I$, satisfying $K^{\star}=-K$, an advanced Green's function for the massless case satisfying $A_{0}^{\star}A_{0} = \rho aI$ and $\overline{A}_{0} = A_{0}$; from this we compute with a mass series expansion $A_{m}$ satisfying $\overline{A}_{m} = A_{m}$ and to a good approximation the quadrature formula.  Furthermore, the radiative Dirac propagator reads $$\Delta_{m} = - A_{m} + A^{\dag}_{m}$$ and we build a theory of fermions in the next section and compute some massless examples in section four in which also different generating structures $K$ are compared.
\section{Dirac Theory.}
Dirac theory contains the equations
$$\{ \psi_{\alpha x},\psi_{\beta y} \} = 0$$
and
$$\{ \psi_{\alpha x}, \overline{\psi}_{\beta y}\}= i\Delta_{\alpha \beta}(x,y)$$
 where in general $\overline{\psi}_{\alpha x} = \psi^{H}_{\beta x} \gamma^{\beta}_{\, \alpha} \otimes 1$ ($H$ denotes the Hermitian conjugate) so that
 $$\{ \psi_{\alpha x}, \psi^{H}_{\beta y} \}= i\Delta_{\alpha \kappa}(x,y)\gamma^{\kappa}_{\beta}.$$ 
Using the spectral decomposition of $i \Delta$, assuming that the latter exists, this equation is equivalent to
$$\{ \psi_{\alpha x}, \psi^{H}_{\beta y} \} = \sum_{r=1}^{k} \lambda_r v_{r,\alpha x} v^{H}_{r,\beta y}  + \sum_{r=1}^{s}\mu_r w_{r, \alpha x} w^{H}_{r, \beta y} + \sum_{r=1}^{t}
	\left( \kappa_r m_{r,\alpha x} n^{H}_{r,\beta y} + \overline{\kappa}_{r} n_{r, \alpha x} m^H_{r, \beta y}  \right)$$ where the orthogonality is with respect $v^{H}\gamma \otimes 1 w$, $\lambda_r,\mu_r \in \mathbb{R}$ and $\kappa_r \in \mathbb{C}$ and $k+s+2t=nm$.  Therefore $i \Delta$ splits representation space $\mathcal{K}  = \mathbb{C}^{m} \otimes \mathbb{C}^{n}$ in three subspaces $\mathcal{K}_{+} \oplus \mathcal{K}_{-} \oplus \mathcal{K}_{0}$ where $\mathcal{K}_{+} = \textrm{Span}_{\mathbb{C}}\{ v_r |r = 1 \ldots k \}$, $\mathcal{K}_{-} = \textrm{Span}_{\mathbb{C}}\{ w_r |r = 1 \ldots s \}$ and $\mathcal{K}_{0} = \textrm{Span}_{\mathbb{C}}\{ m_r, n_r |r = 1 \ldots t \}$.  Define then $\mathcal{H}_{0}$ as the subspace spanned by all eigenvectors or null eigenpairs corresponding to a zero eigenvalue.  Then, define for any $j$ such that $\lambda_j \neq 0$, $a_j= \overline{v}_j \psi$, in case $\mu_j \neq 0$ we have that $b^{\dag}_j = - \overline{w}_j \psi$ and finally in case $\kappa_j \neq 0$ we define $c_j = \overline{n}_j \psi$ and $d_j = \overline{m}_j \psi$.  
In other words, we pose that 
	$$\psi = \sum_{j : \lambda_j \neq 0}v_j a_j  + \sum_{j:\mu_j \neq 0}w_jb_j^{\dag} + \sum_{j:\kappa_j \neq 0}\left( m_jc_j + n_jd_j \right)$$ then\footnote{The reason why we put $b^{\dag}$ rather than $b$ may be found in the calculation of $\overline{\psi}\psi$.  The latter equals $\sum_i a^{\dag}_i a_i - b_i b^{\dag}_i + d^{\dag}_i c_i + c^{\dag}_i d_i$ and demanding that the normal form of this expression comes with coefficients $+1$ is equivalent to putting $b^{\dag}$ \emph{and} using Fermi statistics instead of Bose.   Therefore also the reason for the anticommutator instead of the commutator is to be found here; this is the causal set replacement for the spin statistics theorem.} the above anticommutation relations lead to the following nonzero anticommutation relations (all other relations vanish)
	\begin{eqnarray*}
	\{a_i, a^{\dag}_j \} & = & \lambda_j \delta_{ij} \\*
	\{b_i, b^{\dag}_j \} & = & \mu_j \delta_{ij} \\*
	\{c_i, d^{\dag}_j \} & = & \kappa_j \delta_{ij}
	\end{eqnarray*} as can be computed directly from their definition.  For example, an elementary computation yields $\{ a_j , b_{k}\} = - i\overline{v}_j\Delta w_{k} = 0$ and $\{ a_j , a^{\dag}_{k}\} = i \overline{v}_{j} \Delta v_{k} = \lambda_j \delta_{jk}$.  Now, we are ready to pose the equivalent of the Dirac equations of motion: any $v \in \mathcal{H}_{0}$ automatically satisfying $\overline{v}\Delta = 0$ gives $\overline{v}\psi = 0$.  In short, our theory contains positive and negative norm particles and anti-particles depending whether $\lambda_j, \mu_j > 0$ or smaller than zero.  Also, it contains complex ghosts which cannot be ignored from the prescription; the vacuum $| 0 \rangle$ is then defined as the state annihilated by all $a_i, b_j,c_k$ and $d_k$.  It is rather interesting that negative norm particles enter the prescription of causal set fermions which is no surprise to this author as its use in physics has been advocated in plenty of other places. \\* \\*  In case no spectral decomposition exists due to a degenerate eigenvalue, one has to look for different generating structures since no physical interpretation can be set up in that case.  
	\section{Examples.}    
	We work out massless theory on a diamond with a generating structure given by
	$$K = \frac{1}{2}\left( \begin{array}{cccc} \beta + \delta & 2 \beta & - 2\delta & 0 \\*
	0 & \beta + \delta & 0 & -2\delta \\* 0 & 0 & \beta + \delta & 2\beta \\* 
	0& 0 & 0 & \beta+\delta  \end{array}\right) $$ where, as before, $\beta^2 = -\delta^2=1$ and $\beta \delta + \delta \beta = 0$.  It is easy to verify that $K^2=I$ and $K^{\star}=-K$.  With $a=1=\rho$, the Pauli-Jordan function equals 
	$$i\Delta_{0} = \left( \begin{array}{cccc} \beta+\delta & 
	\beta & - \delta & 0 \\*
	\beta & \beta+ \delta & 0 & -\delta \\* -\delta & 0 & \beta+\delta & \beta \\* 
	0& -\delta & \beta & \beta+\delta \end{array}\right)$$ and $i\Delta_{0}$ and we leave it up to the reader to determine the particle content in the standard two dimensional representation given by $$\beta = \left( \begin{array}{cc}  0 & 1 \\*
	1 & 0 \end{array}\right) $$ and $$\delta =  \left( \begin{array}{cc}  i & 0 \\*
	0 & - i \end{array}\right). $$   We study two distinct generating structures for the causet given by $$I = \left( \begin{array}{cc} 0 & 1 \\*
	0& 0  \end{array}\right)$$ whose particle content is identical. 
	Here the ultrastrong generating structures $$K = \frac{1}{2}\left( \begin{array}{cc} \beta + \delta & 2\beta \\*
	0 & \beta + \delta \end{array}\right)$$ and $$L = \frac{1}{2}\left( \begin{array}{cc} \beta + \delta & -2\delta \\*
	0 & \beta + \delta \end{array}\right)$$ lead to the following Pauli Jordan matrices
	$$i \Delta_{0} = \left( \begin{array}{cc} \beta + \delta & \beta \\*
	  \beta & \beta + \delta \end{array}\right)$$ and $$i \Delta_{0} = \left( \begin{array}{cc} \beta+ \delta & - \delta \\*
	 -\delta &  \beta + \delta \end{array}\right)$$ respectively.  As the reader can easily verify, the first matrix has eigenvalues $i,-i, \pm \sqrt{3}$ whose eigenvectors form a conjugated null pair and one of negative and positive norm leading to a theory with two ghosts and two particles of \emph{positive} norm.  The second matrix has eigenvalues $1,-1, \pm \sqrt{3}i$  whose eigenvectors form a conjugate null pair and one of positive norm and negative norm leading to a theory with two positive norm particles and two ghosts. 
	
\end{document}